\documentclass[final,3p,times,twocolumn,number,sort&compress]{elsarticle}

\usepackage{graphicx}

\usepackage{amssymb}

\renewcommand\({\left(}
\renewcommand\){\right)}
\renewcommand\[{\left[}
\renewcommand\]{\right]}

\newcommand\n{{\mbox {\boldmath $\nabla$}}}
\newcommand{\ra}{\rightarrow}

\def\lsim{\raise 0.4ex\hbox{$<$}\kern -0.8em\lower 0.62
ex\hbox{$\sim$}}

\def\gsim{\raise 0.4ex\hbox{$>$}\kern -0.7em\lower 0.62
ex\hbox{$\sim$}}

\def\lbar{{\hbox{$\lambda$}\kern -0.7em\raise 0.6ex
\hbox{$-$}}}

\newcommand\eq[1]{eq.~(\ref{#1})}

\newcommand\pa{\partial}
\newcommand\p{\partial}

\newcommand\ee{\end{equation}}
\newcommand\be{\begin{equation}}
\def\bea{\begin{array}}
\def\eea{\end{array}}\def\ea{\end{array}}
\newcommand\ees{\end{eqnarray}}
\newcommand\bees{\begin{eqnarray}}

\def\d{\delta}

\def\eps{\epsilon}

\def\dslash{\hspace{-1mm}\not{\hbox{\kern-2pt $\partial$}}}
\def\Dslash{\not{\hbox{\kern-4pt $D$}}}
\def\pslash{\not{\hbox{\kern-2.1pt $p$}}}
\def\kslash{\not{\hbox{\kern-2.3pt $k$}}}
\def\qslash{\not{\hbox{\kern-2.3pt $q$}}}

\newcommand{\vac}{|0\rangle}
\newcommand{\cav}{\langle 0|}

\def\p1{{\bf p}_1}
\def\p2{{\bf p}_2}
\def\k1{{\bf k}_1}
\def\k2{{\bf k}_2}

\newcommand{\emn}{\eta_{\mu\nu}}

\newcommand{\gmn}{g_{\mu\nu}}

\newcommand{\Rmn}{R_{\mu\nu}}

\newcommand{\Tmn}{T_{\mu\nu}}

\newcommand{\dddM}{\kern 0.2em \raise 1.9ex\hbox{$...$}\kern -1.0em \hbox{$M$}}
\newcommand{\dddQ}{\kern 0.2em \raise 1.9ex\hbox{$...$}\kern -1.0em \hbox{$Q$}}
\newcommand{\dddI}{\kern 0.2em \raise 1.9ex\hbox{$...$}\kern -1.0em\hbox{$I$}}
\newcommand{\dddJ}{\kern 0.2em \raise 1.9ex\hbox{$...$}\kern-1.0em
\hbox{$J$}}
\newcommand{\dddcalJ}{\kern 0.2em \raise 1.9ex\hbox{$...$}\kern-1.0em
\hbox{${\cal J}$}}

\newcommand{\dddO}{\kern 0.2em \raise 1.9ex\hbox{$...$}\kern -1.0em
\hbox{${\cal O}$}}
\def\dddz{\raise 1.5ex\hbox{$...$}\kern -0.8em \hbox{$z$}}
\def\dddd{\raise 1.8ex\hbox{$...$}\kern -0.8em \hbox{$d$}}
\def\dddbd{\raise 1.8ex\hbox{$...$}\kern -0.8em \hbox{${\bf d}$}}
\def\ddbd{\raise 1.8ex\hbox{$..$}\kern -0.8em \hbox{${\bf d}$}}
\def\dddx{\raise 1.6ex\hbox{$...$}\kern -0.8em \hbox{$x$}}

\newcommand{\Sch}{Schwarzschild }

\newcommand{\mpl}{M_{\rm Pl}}

\newcommand{\lpl}{l_{\rm Pl}}

\newcommand{\rde}{\rho_{\rm DE}}
\newcommand{\lc}{\Lambda_c}

\journal{Physics Letters B}

\begin{document}

\begin{frontmatter}



\title{Early dark energy from zero-point quantum fluctuations}

\author{Michele Maggiore, Lukas Hollenstein, Maud Jaccard and Ermis Mitsou}
\address{D\'epartement de Physique Th\'eorique and Center for Astroparticle Physics,\\
Universit\'e de Gen\`eve, 24 quai Ansermet, CH-1211 Gen\`eve 4}


\begin{abstract}

We examine a cosmological model with a dark energy density of the form $\rde(t)=\rho_X(t)+\rho_Z(t)$, where $\rho_X$ is the component that accelerates the Hubble expansion at late times and $\rho_Z(t)$ is an extra contribution proportional to $H^2(t)$. This form of $\rho_Z(t)$ follows  from the recent proposal that the contribution of zero-point fluctuations of quantum fields to the total energy density should be computed by subtracting the Minkowski-space result from that computed in the FRW space-time. We discuss theoretical arguments that support this subtraction.
By definition, this eliminates the quartic divergence in the vacuum energy density responsible for the cosmological constant problem. We show that the remaining quadratic divergence can be reabsorbed into a redefinition of Newton's constant only under the assumption that  $\n^{\mu}\cav\Tmn\vac=0$, i.e. that the energy-momentum tensor of vacuum fluctuations is conserved in isolation. However in the presence of an ultra-light scalar field $X$ with $m_X<H_0$, as typical of some dark energy models, the gravity effective action  depends both on the gravitational field and on the $X$ field. In this case
general covariance only requires $\n^{\mu}(T^X_{\mu\nu}+\cav\Tmn\vac)$.
If there is an exchange of energy between these two terms, there are potentially observable consequences. 
We construct an explicit model with an interaction between $\rho_X$ and $\rho_Z$
and we show that the total dark energy density $\rde(t)=\rho_X(t)+\rho_Z(t)$ always remains a finite fraction of the critical density at any time, providing a specific model of early dark energy. We discuss the implication of this result for the coincidence problem and we estimate the model parameters by means of a full likelihood analysis using current CMB, SNe Ia and BAO data.

\end{abstract}

\begin{keyword} early dark energy \sep cosmological constant \sep vacuum fluctuations


\end{keyword}

\end{frontmatter}



\section{Introduction}
\label{sect:intro}

Understanding the origin of dark energy is one of the most important challenges facing cosmology and theoretical physics (see e.g.~\cite{Weinberg:1988cp,Peebles:2002gy,Padmanabhan:2002ji,Copeland:2006wr}). One aspect of the problem is to understand what is the role  of zero-point vacuum fluctuations in cosmology. 
In a  Friedmann-Robertson-Walker (FRW) metric with Hubble parameter $H(t)$  the bare vacuum energy density takes the form
\be\label{c1c2}
\hspace*{-5mm}[\rho_{\rm bare} (\lc)]_{\rm FRW} =[\rho_{\rm bare} (\lc)]_{\rm Mink} 
+{\cal O}\(H^2(t)\lc^2\)\, ,
\ee
where $[\rho_{\rm bare} (\lc)]_{\rm Mink}$ is the bare vacuum energy density in Minkowski space, whose leading divergence is ${\cal O}(\lc^4)$, and we used for definiteness a momentum space cutoff
$\lc$. In the usual treatment this $\lc^4$ divergence is reabsorbed into a renormalization of the cosmological constant, giving rise to the cosmological constant problem. The divergence $\propto H^2\lc^2$ is instead absorbed into a renormalization of Newton's constant $G$~\cite{Fulling:1974zr,Birrell:1982ix}.

In this paper, expanding on results presented in
\cite{Maggiore:2010wr}, we reexamine the role of vacuum energies in cosmology. First, we will propose theoretical arguments suggesting that
the correct way of computing  the physical vacuum energy is to subtract the bare vacuum energy density of Minkowski space, $[\rho_{\rm bare} (\lc)]_{\rm Mink}$, from the  FRW result  given in \eq{c1c2}, before renormalizing the result. By definition this subtraction eliminates the troublesome $\lc^4$ divergence and, therefore, the cosmological constant problem. Then we turn our attention to the left over term $H^2\lc^2$ which now becomes the leading term in the vacuum energy. 
It is usually believed that this quadratic divergence can be reabsorbed into a renormalization of $G$. We show that this is correct 
only under the assumption that vacuum expectation value (VEV) of the energy-momentum tensor is conserved in isolation (an assumption that was implicit in the literature). However, general covariance of General Relativity (GR) only implies the conservation of the {\em total} energy-momentum tensor $\Tmn+\cav\Tmn\vac$, including both  the classical term $\Tmn$ and the semiclassical term $\cav\Tmn\vac$. The separate conservation of
$\cav\Tmn\vac$ only takes place if we can define an effective  action which depends only on the gravitational field, by integrating out the matter degrees of freedom. This is possible only if the matter degrees of freedom are heavy with respect to the energy scale of the problem, and can then be integrated out. In a cosmological setting, this means that matter fields should satisfy $m>H_0$. If, in contrast, there is an ultra-light scalar field with $m<H_0$, as is typical for dark energy models such as quintessence, this field cannot be integrated out from the effective low energy action. 
We show that, as a result, in general $\n^{\mu}T^X_{\mu\nu}=-\n^{\mu}\cav\Tmn\vac\neq 0$.
In this case
the effect of the quadratically divergent term in the vacuum fluctuations cannot simply be absorbed into a renormalization of Newton's constant $G$, and gives rise to interesting and potentially detectable cosmological effects. We construct a specific coupled early dark energy model and test it against current observations.

We use natural units where $\hbar=c=1$, $G=\mpl^{-2}$. If not specified otherwise, we work in a spatially flat FRW metric with signature ($-$+++), cosmic time $t$, scale factor $a(t)$ and Hubble parameter $H(t)=(da/dt)/a$. Today, the Hubble parameter and the critical density take the values $H_0$ and $\rho_0=3H_0^2/(8\pi G)$, respectively.

\section{Subtraction of the flat-space vacuum energy}

In Minkowski space the divergence in the vacuum energy density is usually dealt with by normal ordering the Hamiltonian, which gives by definition a vanishing result for the physical vacuum energy density.
However, it is useful to realize that the problem can be treated more generally in the context of renormalization theory, which rather allows us to fix the renormalized vacuum energy density to any observed value. In the standard language of renormalization, divergences in a generic $N$-point Green's function are cured by adding the corresponding counterterms to the Lagrangian density. The same procedure can be applied to vacuum energy, i.e.~to   the $N=0$ Green's function: one simply adds a constant counterterm $-\rho_{\rm count}(\lc)$ to the Lagrangian density. This corresponds to adding a term $+\rho_{\rm count}(\lc)$ to the Hamiltonian density. Hence the renormalized, physical vacuum energy density is given by 
$\rho_{\rm ren}=\rho_{\rm bare}(\lc)+\rho_{\rm count}(\lc)$. As always in renormalization theory, the counterterm $\rho_{\rm count}$ is chosen so to cancel the divergences in $\rho_{\rm bare}$ and leave us with the desired finite part that is fixed by comparison with the experiment.

Using the language of renormalization theory is useful in this context because it makes clear that the cosmological constant problem is not that
quantum field theory (QFT) gives a {\em wrong} prediction for the cosmological constant
(as it is sometimes incorrectly said). Strictly speaking QFT makes no prediction for the cosmological constant, just as it does not predict the electron mass nor the fine structure constant. Rather, it is a problem of naturalness, in the sense that the counterterm 
$\rho_{\rm count} (\lc)$ must be fine-tuned to  exceeding accuracy, in order to cancel the $\lc^4$ divergence in $\rho_{\rm bare}$, leaving a physical vacuum energy density that, if one identifies $\lc$ with the Planck mass, is about
${\cal O}(10^{120})$ times smaller than $\lc^4$.

Posing the problem in terms of a cancellation between $\rho_{\rm bare}(\lc)$  and $\rho_{\rm count}(\lc)$
can also give a first hint for a possible solution. First of all, one should appreciate that  neither the bare vacuum energy $\rho_{\rm bare}(\lc)$ 
nor the counterterm $\rho_{\rm count}(\lc)$ have a physical meaning and only their sum is an observable. Thus, this kind of cancellation is different from a fine-tuning between observable quantities.
Indeed, the Casimir effect is a well-known example where a rather similar cancellation takes place. In that case the physical vacuum energy density of a quantum field in a finite volume is found by taking the difference between the bare vacuum energy density computed in this finite volume and
the bare vacuum energy density in an infinite volume. Regularizing with a cutoff $\lc$  both terms diverge as $\lc^4$, but their difference is finite and depends only on the physical size of the system.  
This might suggest that, similarly,
to obtain the physical effect of the vacuum energy density in cosmology, one should  compute the vacuum energy density in a FRW space-time and subtract from it the value computed in a reference geometry, which could be naturally taken as Minkowski space, leading to a sort of  ``cosmological Casimir effect". 

Before taking this analogy with the Casimir effect seriously, one must however face the obvious objection that in special relativity the zero of the energy can be chosen arbitrarily, and only energy differences with respect to the ground state are relevant.\footnote{Equivalently, one may observe that in the Casimir effect  one actually measures the force between the plates, i.e.~not the energy density itself but only its derivative w.r.t.~the size of the system $L$. Since the divergence $\lc^4$ is independent of $L$, it can simply be dropped.}
In contrast, in GR we cannot chose the zero of the energy arbitrarily. One typically expects that ``every form of energy gravitates", so the contribution of Minkowski space cannot just be dropped.

While it is certainly true that in GR the choice of the zero for the energy is not arbitrary, the point that we wish to make here is that what is the correct choice  can be a non-trivial issue.
As a  first example, consider  the definition of energy for asymptotically flat space-times. This is 
obtained from the
Hamiltonian formulation of GR, which goes back to the classic paper by Arnowitt, Deser and Misner (ADM) \cite{Arnowitt:1962hi,pois04}. To properly define the Hamiltonian of a given field configuration in GR one must work  at first in a finite three-dimensional volume $V$. Then the Hamiltonian  takes the form $H_{\rm GR}=H_{\rm bulk}+H_{\rm boundary}$, where $H_{\rm bulk}$ is given by an integral over the spatial volume $V$ at fixed time, while  $H_{\rm boundary}$
is given by an  integral over the  two-dimensional boundary $\pa V$. 
When one evaluates $H_{\rm bulk}$ on any classical solution of the equations of motion one finds a vanishing result (since $H_{\rm bulk}$ is proportional to the constraint equations of GR), so the whole contribution comes from the boundary term. On the other hand,
$H_{\rm boundary}$ diverges for any asymptotically flat metric $\gmn$ (including flat space-time),
when the boundary is finally taken to infinity. The solution proposed by ADM is to subtract from this boundary term the same term computed in Minkowski space $\emn$. Accordingly, the energy $E$ associated with a classical asymptotically flat metric $\gmn$ is obtained by defining
\be\label{EHH}
E=H_{\rm GR}[\gmn]-H_{\rm GR}[\emn]\, .
\ee
This provides the standard definition of mass in GR, and reproduces the expected properties of asymptotically flat space-times. For instance, when applied to the \Sch space-time, it correctly gives the mass  that appears in the \Sch metric. This  underlines that our intuition that any form of energy gravitates according to GR is not entirely correct: \eq{EHH} tells us that the energy associated to Minkowski space does not gravitate.

Similar subtractions also hold for non-asymptotically flat space-times, and can be performed either by subtracting the contribution of some reference space-time whose boundary has the same induced metric as the background under consideration~\cite{Gibbons:1976ue,Brown:1992br,Brown:1994gs,Hawking:1995fd}, or even without introducing a reference background but just by adding some local counterterms to the boundary action, given by a coordinate-invariant functional of the intrinsic boundary geometry~\cite{Balasubramanian:1999re,Kraus:1999di}. The latter prescription is particularly appealing for asymptotically AdS space-times. In fact, in the context of the AdS/CFT correspondence, this way of removing divergences in the gravitational action on the AdS side corresponds to the renormalization of the UV divergences in the conformal QFT that lives on the boundary \cite{Balasubramanian:1999re,Kraus:1999di,Lau:1999dp,Mann:1999pc,Emparan:1999pm}.

These examples show that, already in classical GR, the energy that actually acts as a source of gravity can be obtained from a Hamiltonian only after performing an appropriate subtraction. It is quite natural to assume that the same should hold at the quantum level, so in particular for zero-point fluctuations of quantum fields in curved space. To understand what is the appropriate subtraction for the FRW metric, we consider the Friedmann equation that results from the Einstein equations sourced by $T^\mu_\nu+\cav T^\mu_\nu\vac$, where $T^\mu_\nu={\rm diag}(-\rho,p,p,p)$ is the ordinary contribution of matter, radiation, etc, and $\cav T^\mu_\nu\vac={\rm diag}(-\rho_{\rm vac},p_{\rm vac},p_{\rm vac},p_{\rm vac})$ is the corresponding contribution of zero-point fluctuations:
\be
H^2(t)=\frac{8\pi G}{3}\( \rho + \rho_{\rm vac}\)\,.
\ee
We then require that Minkowski space, $H(t)=0$, should be a solution in the limit $\rho\!\ra\! 0$. This implies that all terms in $[\rho_{\rm vac}]_{\rm FRW}$ that do not vanish for $H\!\ra\! 0$ must be subtracted. In other words, we must subtract the vacuum energy computed in Minkowski space.\footnote{This conclusion also fits nicely with that of ref.~\cite{Brustein:2000hh} where the authors considered the QFT of a large number ${\cal N}$ of fields in Minkowski space, and found that the vacuum fluctuations collapse to black holes on scales smaller than ${\cal O}( {\cal N}^{1/4}\lpl)$. They conclude that Minkowski space would therefore be unstable to black hole formation unless either the length-scale where quantum gravity sets in, for a theory with ${\cal N}$ fields, is of order ${\cal N}^{1/4}\lpl$
, or the vacuum fluctuations in Minkowski space do not gravitate.}  In contrast, all terms proportional to $H^2$ or $H^4$ are consistent with this requirement and thus generically allowed. This procedure eliminates the $\lc^4$ term (as well as flat-space terms that appear for massive fields, such as $m^2\lc^2$ and $m^4\ln\lc$, and are also much larger than the observed vacuum energy, for all known massive particles). Thus, this treatment of vacuum fluctuations solves the cosmological constant problem by definition, at least in its ``old" form, i.e.~it explains why the observed vacuum energy is many orders of magnitude smaller than $\mpl^4$.

We next turn to the quadratic divergence left over after the subtraction. We will see that it could teach us something about the more recent form of the cosmological constant problem, namely explaining why the observed dark energy (DE) density is just of the order of the critical density of the universe today, the coincidence problem.

\section{Non-interacting zero-point fluctuations and renormalization of $G$}
\label{sect:renG}

Let us now discuss the fate of the quadratic divergence $\propto\lc^2H^2(t)$ in \eq{c1c2}. To understand what is the structure of the renormalized VEV of the energy-momentum tensor, consider
the semiclassical Einstein equations for the renormalized quantities,
\be
\Rmn-\frac{1}{2}\gmn R=8\pi G (\Tmn+\cav\Tmn\vac)\, ,
\ee
in which the vacuum expectation value of $\Tmn$ is added as an additional source term.
Together with the Bianchi identities the above equation implies
\be\label{totcons}
\n^{\mu} (\Tmn+\cav\Tmn\vac)=0\, .
\ee
This equation is therefore a consequence of the general covariance of the renormalized theory.\footnote{Observe that general covariance holds, for  renormalized quantities, independently of the regularization scheme employed, as it should. If one uses a scheme that breaks general covariance at the level of regularization,  such as a momentum space cutoff, then to recover it at the level of renormalized quantity one must also add non-covariant counterterms. See~\cite{hjmm} for an extended discussion of this point, and comparison with the literature.} 
If $\Tmn$ and $\cav\Tmn\vac$ are separately conserved we further have 
$\n^{\mu}\Tmn=\n^{\mu}\cav\Tmn\vac=0$. 

The stronger condition $\n^{\mu}\cav\Tmn\vac=0$ can  indeed be derived by using
the effective action for gravity, which is
obtained by treating the metric $\gmn$ as a classical background and integrating out the matter degrees of freedom
(see e.g. refs.~\cite{Barvinsky:1985an,Buchbinder:1992rb,Shapiro:2008sf}). The   VEV of the renormalized energy-momentum tensor  is then obtained by taking  the functional derivative
$(2/\sqrt{-g})\d/\d\gmn$ of the effective action. One can perform the calculation of the effective action using regularizations, such as dimensional regularization or point-splitting, that preserve general covariance explicitly. The effective action is therefore explicitly generally covariant, and the VEV of the energy-momentum tensor derived from it is automatically covariantly conserved. In this case,  all divergences in $\cav\Tmn\vac$ can be reabsorbed into generally covariant counterterms in the effective action.  
In particular, the divergence $\propto H^2(t)\lc^2$ is cured by a counterterm proportional to the Einstein-Hilbert action. In fact, taking the variation of the Einstein-Hilbert term gives the Einstein tensor $G_{\mu\nu}$, and $G_{00}$, specialized to the FRW metric, is proportional to $H^2(t)$. 
This shows that a VEV $\cav T_{00}\vac\propto H^2(t)$ is obtained, in the effective action language, by an additional term proportional to 
$\int d^4x\sqrt{-g}\, R$, and therefore is reabsorbed into a renormalization of Newton's constant, as it  is well known~\cite{Fulling:1974zr,Birrell:1982ix}.

Such an effective action approach, however, assumes that we can integrate out all matter fields, i.e.~that they are sufficiently massive with respect to the scale of interest. In our cosmological context this means that we are implicitly assuming that all fields have a mass $m$ bigger than the Hubble parameter $H(t)$  at the time of interest, which is the quantity that fixes the relevant scale. Equivalently, we are assuming that the wavelength $1/m$ is smaller than the horizon size $H^{-1}(t)$. It is however  interesting to consider the case in which in the spectrum there is a scalar particle $X$ with a mass $m_X<H_0$. Such an ultra-light scalar field, with the mass protected against radiative correction by demanding that it is a pseudo Nambu-Goldstone boson, provides in fact a typical realization of quintessence~\cite{Frieman:1995pm}. In this case we cannot integrate out this field, and the low-energy effective action  necessarily depends both on the metric and on this scalar field. Then  the above derivation giving
$\n^{\mu} \cav\Tmn\vac=0$ no longer goes through. By taking the variation of this action  with respect to the metric we get the total energy-momentum tensor, including both
$\cav\Tmn\vac$ and the energy-momentum tensor $T^X_{\mu\nu}$ of this scalar field, and general covariance now only implies $\n^{\mu}( \cav\Tmn\vac + T^X_{\mu\nu})=0$.

Whether these two terms are separately conserved is now a dynamical question, and depends on whether there is an interaction among them.
One can imagine mechanisms by which vacuum fluctuations can exchange energy with other forms of matter. Typical examples are the amplification of vacuum fluctuations \cite{Grishchuk:1974ny,Starobinsky:1979ty}, or the change in a large-scale scalar field due to the continuous  horizon-crossing of small-scale quantum fluctuations of the same scalar field, which is also at the basis of stochastic inflation~\cite{Starobinsky:1986fx}.
If $\n^{\mu} \cav\Tmn\vac =-\n^{\mu} T^X_{\mu\nu}\neq 0$, it is no longer possible to reabsorbe
the effect of $ \cav\Tmn\vac$  into a renormalization of the Einstein-Hilbert term, nor of any other generally covariant local operator in the effective action. In fact, taking the functional derivative
$(2/\sqrt{-g})\d/\d\gmn$ of a generally covariant term, we necessarily obtain a covariantly conserved 
tensor, so we can never obtain a quantity $ \cav\Tmn\vac$ that satisfies
$\n^{\mu} \cav\Tmn\vac\neq 0$.

\section{Cosmology with zero-point fluctuations}

We now explore the cosmological consequences of the hypothesis that $\cav\Tmn\vac$ and $\Tmn$ are  conserved in conjunction, as in \eq{totcons}, but not separately. In this case, as shown above, the term $\lc^2H^2$ cannot be reabsorbed into a renormalization of $G$, and we rather expect that it will give a genuine
contribution to the total energy density 
$\rho_Z(t)= {\cal O}(H^2(t)M^2)$,
where $M$ is the UV scale where new physics sets in
(so that $M$ could be typically given by the Planck mass $\mpl$, or by the string mass). Recalling that 
the critical density  is 
$\rho_c(t)=3H^2(t)\mpl^2/(8\pi)$ we see that, for $M$ of order $\mpl$, $\rho_Z(t)$ is of order of the critical density $\rho_c(t)$ at any time $t$. 
Thus we write
\be\label{rhoZrhoc}
\rho_Z(t)=\Omega_Z\rho_c(t)=\Omega_Z\rho_0H^2(t)/H_0^2\, ,
\ee
where $\rho_0$ is the present value of the critical density. The value of $\Omega_Z$ is fixed by the renormalization procedure to the observed value, so we keep it as a free parameter. The same is true for the equation of state (EOS) parameter $w_Z$ defined by $p_Z=w_Z\rho_Z$, that can in principle be a function of time (see \cite{hjmm} for an extended discussion of this point).

The late time acceleration of the Universe cannot be explained only by a DE density that scales 
like $H^2(t)$~\cite{Basilakos:2009wi,Maggiore:2010wr,Grande:2011xf}, basically because observations tell us that the total DE density is at least approximately constant in the recent cosmological epoch. Therefore we assume $\rho_Z(t)$ only to provide a part of the total DE density that we write as $\rde(t)=\rho_X(t)+\rho_Z(t)$. Here $\rho_X(t)$ is a second dynamical DE component that is the one dominating the energy budget at the current epoch. In this approach the physical origin of $\rho_X$ and $\rho_Z$ can a priori be completely different. For instance, $\rho_X$ could be due to a scalar field, as in quintessence models. 
In particular, we will take $\rho_X$ to be due to an ultra-light scalar field with $m<H_0$  as discussed in sect.~\ref{sect:renG}.
We define the EOS parameter of the $X$-component by $w_X=p_X/\rho_X$. Since  $\rho_X$ and $\rho_Z$  have different origins, $w_X$ and $w_Z$ can in principle be different and we will see that only for $w_X\neq w_Z$ a tracking mechanism for DE emerges.

Mechanisms such as the amplification of vacuum fluctuations can produce an interaction between $\rho_Z$ and scalar fields, but do not lead to interactions with photons nor massless fermions, that are not amplified in the FRW space-time because of conformal invariance. As we saw in sect.~\ref{sect:renG},
massive particles with $m>H_0$ can be integrated out in the effective action and therefore cannot contribute to the violation of the condition $\n^{\mu} \cav\Tmn\vac=0$.
Thus, in our context it is  natural to consider an interaction between $\rho_Z$ and $\rho_X$, while non-relativistic matter and radiation are conserved in isolation, so they scale in the standard way, $\rho_M\sim a^{-3}$ and $\rho_R\sim a^{-4}$. In contrast, $\rho_Z$ and $\rho_X$ satisfy the coupled conservation equation
\be\label{consrl}
\dot{\rho}_Z+\dot{\rho}_{X}+3(1+w_Z)H\rho_Z+3(1+w_{X})H\rho_{X}=0\, .
\ee
In terms of the total DE density, $\rde(t)= \rho_{X}(t)+\rho_Z(t)$, it reads
\be\label{consDE}
\dot{\rho}_{\rm DE}+3(1+w_{X})H\rde=3(w_{X}-w_Z)H\rho_Z\, .
\ee
We insert $\rho_Z(t)=\Omega_Z\rho_0H^2/H_0^2$ on the r.h.s.~and use the Friedmann equation, that now reads $H^2(t)/H_0^2=\Omega_R a^{-4}+\Omega_M a^{-3}+\rde(t)/\rho_0$. With the definitions
\be
w_0\equiv w_X+\eps\, ,\qquad
\eps\equiv \Omega_Z(w_Z-w_X)\, ,
\ee
and using $d/dt =a H (d/da)$, we  obtain
\be\label{eq:rhoZrhoLfull}
\[a\frac{d}{da}+3(1+w_0)\]\frac{\rde}{\rho_0}=
-3\eps
\[ \frac{\Omega_R}{a^4}+\frac{\Omega_M}{a^3}\]\, .
\ee
Note that the background evolution of the total DE density is fully determined by the two parameters $w_0$ and $\epsilon$ which can in principle be functions of time. We assume these functions to be constant for the scope of this work, i.e.~we assume $w_Z$ and $w_X$ to be constant. A more general analysis will be presented in \cite{hjmm}. Then the evolution can be solved analytically,
\be\label{evolrhoDE}
\frac{\rde}{\rho_0}= \frac{C} {a^{3(1+w_0)}}
+\eps
\[  \frac{1}{\frac{1}{3}-w_0}\, \frac{\Omega_R}{a^4}-
\frac{1}{w_0}\,
\frac{\Omega_M}{a^3}\]
\, ,
\ee
where $C$ is the integration constant that is fixed by the condition $\rde(a=1)/\rho_0=\Omega_{\rm DE}=1-(\Omega_R+\Omega_M)$. Observe that the model reduces to $w$CDM when $w_X=w_Z$ and the deviation from $w$CDM only depends on $\eps$, so we are dealing with a one-parameter extension of $w$CDM, that we will call $w$ZCDM.

\begin{figure}
\includegraphics[width=0.88\columnwidth]{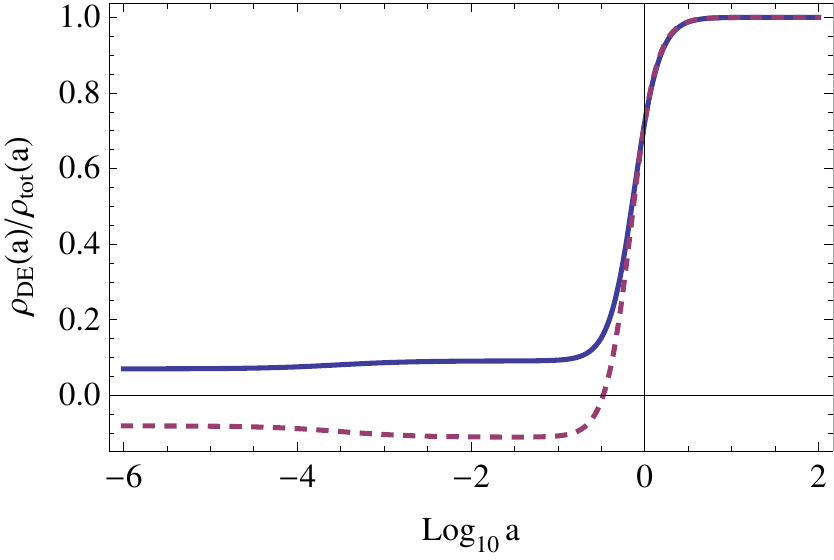}\\
\caption{\label{fig:tracking}
The energy fraction $\rde(a)/\rho_{\rm tot}(a)$, for $w_0=-1$ and $\eps =0.1$ (solid line) and $\eps =-0.1$ (dashed).}
\end{figure}

\begin{figure}
\includegraphics[width=0.9\columnwidth]{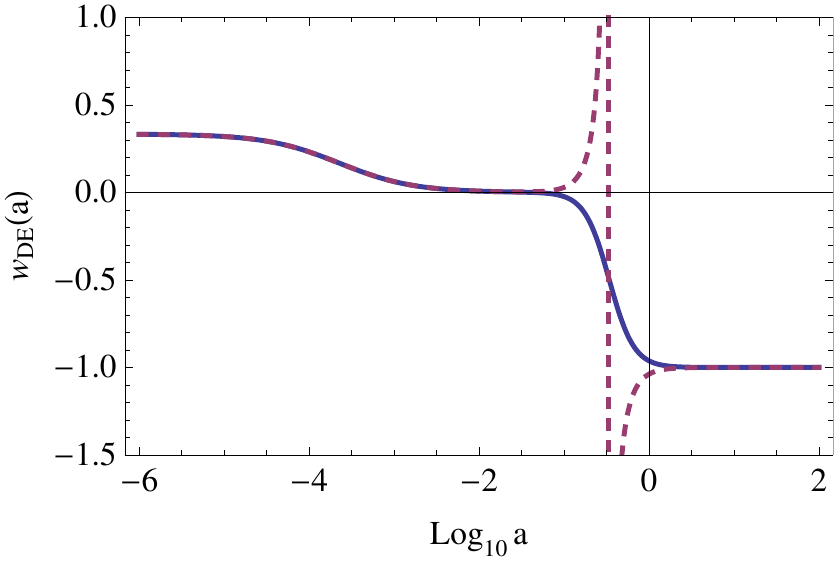}
\caption{\label{fig:wDE}
The equation of state parameter $w_{\rm DE}(a)$ for the same models as in fig.~\ref {fig:tracking}.}
\end{figure}

A very interesting feature of (\ref{evolrhoDE}) is that $\rde$ always scales as the dominant energy component, for $\eps\neq 0$. In  fig.~\ref{fig:tracking} we show the ratio $\rde(a)/\rho_{\rm tot}(a)$, where $\rho_{\rm tot}(a)=\rde(a)+\rho_0\Omega_Ma^{-3}+\rho_0\Omega_R a^{-4}$. Deep into the RD phase, as well as in the MD phase, $\rde(a)/\rho_{\rm tot}(a)={\cal O}(\eps)$, while today it becomes ${\cal O}(1)$. Compared to $\Lambda$CDM, where the ratio $\rho_{\Lambda}/\rho_{\rm tot}$ is of order one today, but goes to zero as $a^3$ during MD and as $a^4$ during RD, the coincidence problem is sensibly alleviated.

We cannot claim that $w$ZCDM  entirely solves the coincidence problem; in this model, in fact, the transition between the regime where $\rde/\rho_{\rm tot}={\cal O}(\eps)$ to the regime where $\rde/\rho_{\rm tot}={\cal O}(1)$ takes place at the present epoch simply because we have fixed the integration constant $C$ in (\ref{evolrhoDE}) by the requirement that $\rde(a=1)/\rho_0=\Omega_{\rm DE}$. Nevertheless the coincidence  problem is certainly alleviated, compared to $\Lambda$CDM, where $\rde$ is parametrically different from $\rho_{\rm tot}$, and the ratio $\rde/\rho_{\rm tot}$ evolves from ${\cal O}(10^{-120})$ at a Planck time to $\sim 0.7$ today.

It is also interesting to note that our model provides a different theoretical justification for 
parameterizations of early DE models that have been proposed in the literature. To make the relation explicit, it is useful to define $w_{\rm DE}=p_{\rm DE}/\rde$ and derive it from the total DE continuity equation, $\dot{\rho}_{\rm DE}+3H(\rde+p_{\rm DE})=0$. Inserting the explicit solution for $\rde(t)$ found in (\ref{evolrhoDE}) we get
\be\label{wDE(t)}
w_{\rm DE}(t) = w_0 + \epsilon [\rho_M(t) + \rho_R(t)]/ \rho_{DE}(t) .
\ee
This function is shown in  fig.~\ref{fig:wDE}, setting $w_0=-1$ for definiteness. For $\eps>0$ it evolves smoothly from a value $w_{\rm DE}\simeq 1/3$ during RD, to $w_{\rm DE}\simeq 0$ during MD and finally goes asymptotically to $w_{\rm DE}\simeq w_0$. This EOS is quite similar to that obtained in a commonly used  parameterization of early dark energy \cite{Doran:2006kp,Hollenstein:2009ph}. Note that, for $\eps <0$, $w_{\rm DE}$ goes through infinity, as a consequence of the fact that $\rde$ goes through zero with $\dot{\rho}_{\rm DE}\neq 0$, see fig.~\ref{fig:tracking}, but the pressure $w_{\rm DE}\rde$ stays finite, and the background evolution is regular.

Our model has some similarities, as well as important differences, with other DE models studied in the literature. In particular, in \cite{Shapiro:2000dz,Shapiro:2003ui,EspanaBonet:2003vk,Shapiro:2009dh} a model was proposed where the DE density has the form $\rho_{\Lambda}(t)=n_0+n_1H^2(t)$, inspired by the idea that the cosmological constant could run under renormalization group. In this case, however, the two components $n_0$ and $n_1H^2(t)$ necessarily have the same EOS parameter. More closely related is the $\Lambda$XCDM model proposed in \cite{Grande:2006nn}, in which DE has two components, an energy density $\rho_{\Lambda}=n_0+n_1H^2(t)$ associated to a running of the cosmological constant interacting with an unspecified dynamical ``cosmon'' field, although in our case the interaction is rather between $\rho_X$ (that, for $w_X\simeq -1$, plays basically the role of $n_0$) and $\rho_Z\sim H^2$, which has a different EOS.

\begin{figure}
\centering
\includegraphics[width=\columnwidth]{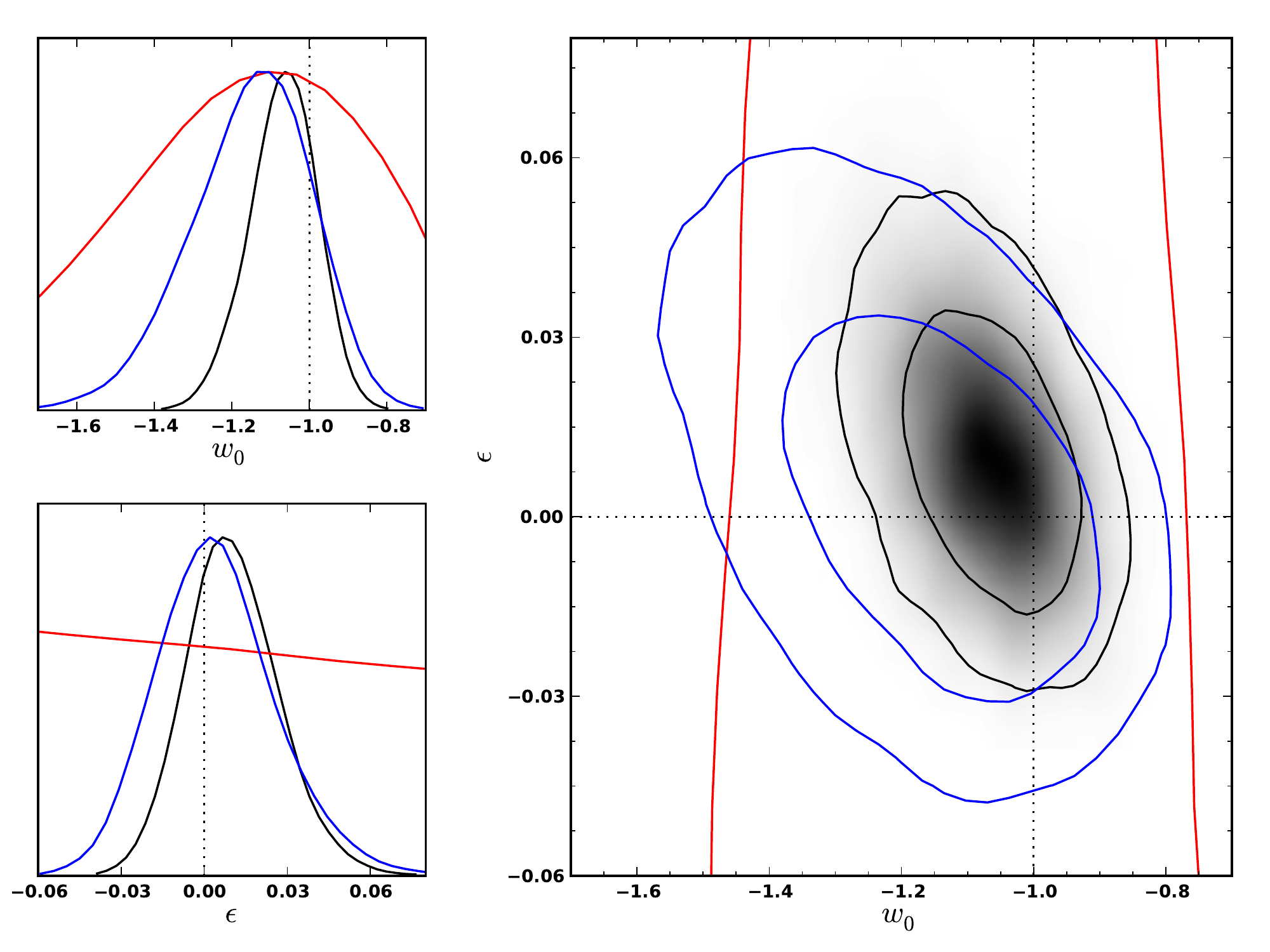}
\caption{\label{fig:ZCSBex_mytri}
Left: The marginalized posterior probabilities of $w_0$ (upper panel) and $\eps$ (lower panel). Right: $1\sigma$ and $2\sigma$ mean-likelihood contours in the $(\eps,w_0)$ plane, after marginalization over all other parameters. Red is SNe only, blue is CMB only, and black is CMB+SNe+BAO.}
\end{figure}

We have performed a detailed comparison of our model with current observations of CMB, SNe Ia and BAO using modified versions of CAMB and CosmoMC \cite{Lewis:1999bs,Lewis:2002ah}, treating perturbations in the DE by modeling it as a perfect fluid (without anisotropic stress) with the EOS parameter $w_{\rm DE}$ given in (\ref{wDE(t)}) and a unit rest-frame sound speed. Full details will be reported in \cite{hjmm}. In fig.~\ref{fig:ZCSBex_mytri} we give a sample of our results. The plots on the left are the one-dimensional posterior probabilities marginalized over all parameters except $w_0$ or $\eps$, respectively, while the plot on the right shows the 2-dimensional 
posterior probability marginalized over all parameters except the pair $(\eps,w_0)$. In particular, we find the marginalized limits $-1.25 < w_0 < -0.908$ and $-0.0201 < \epsilon < 0.0460$ at 95\% C.L., consistent with $\Lambda$CDM. The means of the marginalized posteriors are at $\langle w_0\rangle=-1.07$ and $\langle \epsilon\rangle=0.0104$ and the standard deviations are $\sigma_{w_0}=0.0873$ and $\sigma_\epsilon=0.0167$, respectively. Thus, our $w$ZCDM model is consistent with current data, and its deviations from $w$CDM, expressed by the parameter $\eps$, are constrained at the level ${\cal O}(10^{-2})$. Future data will be able to set more stringent limits or to detect a non-vanishing value of $\eps$.

\vspace{1mm}
{\em Acknowledgments}. We thank R.~Brustein, S.~Foffa, M.~Frommert, R.~K.~Jain, M.~Kunz, M.~Shaposhnikov, R.~Rattazzi and A.~Riotto for very useful discussions. This work is supported by the Fonds National Suisse.


\bibliographystyle{elsarticle/model1a-num-names}
\bibliography{paperplbv2}

\end{document}